\documentclass{ws-mpla}

\begin{document}

\markboth{H.Suganuma}
{Inter-Quark Potentials in Baryons and 
Multi-Quark Systems in QCD}

\catchline{}{}{}{}{}

\title{INTER-QUARK POTENTIALS IN BARYONS AND MULTI-QUARK SYSTEMS IN QCD
}

\author{\footnotesize HIDEO SUGANUMA, ARATA YAMAMOTO, NAOYUKI SAKUMICHI
}

\address{Graduate School of Science, Kyoto University, \\
Kitashirakawaoiwake, Sakyo, Kyoto 606-8502, Japan
\\
suganuma@ruby.scphys.kyoto-u.ac.jp}

\author{TORU T. TAKAHASHI, HIDEAKI IIDA}

\address{Yukawa Institute for Theoretical Physics, Kyoto University, \\
Kitashirakawaoiwake, Sakyo, Kyoto 606-8502, Japan}

\author{FUMIKO OKIHARU}

\address{Faculty of Education and Human Sciences, Niigata University, \\
Ikarashi 2-8050, Niigata 950-2181, Japan}

\maketitle

\pub{Received 17 Feb. 2008}{Revised (Day Month Year)}

\begin{abstract}
We perform the first studies of various inter-quark potentials in SU(3)$_{\rm c}$ lattice QCD.
From the accurate lattice calculation for more than 300 different patterns of three-quark (3Q) systems, 
we find that the static 3Q potential $V_{\rm 3Q}$ is well described by Y-Ansatz, i.e., 
the Coulomb plus Y-type linear potential. 
Quark confinement mechanism in baryons is also investigated in maximally-Abelian projected QCD.
We next study the multi-quark potentials $V_{n{\rm Q}}$ ($n$=4,5) in SU(3)$_{\rm c}$ lattice QCD, 
and find that they are well described by the one-gluon-exchange Coulomb plus multi-Y type linear potential, 
which supports the flux-tube picture even for the multi-quarks. 
Finally, we study the heavy-heavy-light quark (QQq) potential both in lattice QCD and in a lattice-QCD-based quark model.

\keywords{Lattice QCD; Confinement; Multi-quarks.}
\end{abstract}

\ccode{PACS Nos.: 12.38.Gc, 12.38.Aw, 12.39.Mk, 12.39.Pn}

\section{Introduction}

In 1966, Yoichiro Nambu\cite{N66} first proposed the SU(3)$_{\rm c}$ gauge theory, i.e., quantum chromodynamics (QCD),  
as a candidate for the fundamental theory of the strong interaction, 
just after the introduction of the ``new" quantum number, ``color".\cite{HN65}
In 1973, the asymptotic freedom of QCD was theoretically proven,\cite{GWP73}
and, through the applicability check of perturbative QCD to high-energy hadron reactions, 
QCD has been established as the fundamental theory of the strong interaction.
However, in spite of its simple form, QCD creates thousands of hadrons and leads to various interesting nonperturbative phenomena 
such as color confinement\cite{N6970,N74} and dynamical chiral-symmetry breaking.\cite{NJL61}
Even at present, it is very difficult to deal with QCD analytically due to its strong-coupling nature in the infrared region.
Instead, lattice QCD has been applied as a direct numerical analysis for nonperturbative QCD.

In 1979, the first application of lattice QCD Monte Carlo simulations\cite{C7980} was done by M.~Creutz 
for the inter-quark potential between a quark and an antiquark using the Wilson loop.
Since then, the study of inter-quark forces has been one of the important issues in lattice QCD.\cite{R05}
Actually, in hadron physics, the inter-quark force can be regarded as an elementary quantity 
to connect the ``quark world" to the ``hadron world", and plays an important role to hadron properties. 

In this paper, we perform the first detailed studies of 
the various inter-quark forces in the three-quark and the multi-quark systems 
with SU(3)$_{\rm c}$ lattice QCD.\cite{TS0102,OST05}  

\section{The Three-Quark Potential in SU(3) Lattice QCD}

In general, the three-body force is regarded as a residual interaction in most fields in physics.
In QCD, however, the three-body force among three quarks is 
a ``primary" force reflecting the SU(3) gauge symmetry.
In fact, the three-quark (3Q) potential is directly responsible 
for the structure and properties of baryons, 
similar to the relevant role of the Q$\bar{\rm Q}$ potential for meson properties. 
Furthermore, the 3Q potential is the key quantity to clarify the quark confinement in baryons.
However, in contrast to the Q$\bar{\rm Q}$ potential,\cite{R05}
there were almost no lattice QCD studies  
for the 3Q potential before our study in 1999,\cite{TS0102}
in spite of its importance in hadron physics. 

As for the functional form of the inter-quark potential, we note two theoretical arguments 
in short and long distance limits.
\begin{itemlist}
\item[1.]
At short distances, perturbative QCD is applicable,
and the inter-quark potential is expressed as a sum of two-body one-gluon-exchange (OGE) Coulomb potentials. 
\item[2.]
At long distances, the strong-coupling expansion of QCD is plausible, and it 
leads to the flux-tube picture.
\end{itemlist}
Then, we theoretically conjecture the functional form of the inter-quark potential 
as the sum of OGE Coulomb potentials and the linear potential based on the flux-tube picture.
Of course, it is highly nontrivial that these simple arguments on UV and IR limits of QCD hold for the intermediate region. 
Nevertheless, the Q$\bar {\rm Q}$ potential $V_{\rm Q\bar Q}(r)$ is well described with this form 
as\cite{R05,TS0102} 
\begin{eqnarray}
V_{\rm Q \bar Q}(r)=-\frac{A_{\rm Q\bar Q}}{r}+\sigma_{\rm Q \bar Q}r+C_{\rm Q\bar Q}.
\label{VQQ}
\end{eqnarray}
For the 3Q system, there appears a junction which connects the three flux-tubes from the three quarks, 
and Y-type flux-tube formation is expected.\cite{TS0102}
Then, the 3Q potential is expected to be 
the Coulomb plus Y-type linear potential, i.e., Y-Ansatz,
\begin{eqnarray}
V_{\rm 3Q}=-A_{\rm 3Q}\sum_{i<j}\frac1{|{\bf r}_i-{\bf r}_j|}+
\sigma_{\rm 3Q}L_{\rm min}+C_{\rm 3Q},
\label{V3Q}
\end{eqnarray}
where $L_{\rm min}$ is the minimal total length of the Y-shaped flux-tube.

\subsection{Manifestly Gauge-Invariant Lattice QCD Results}

For more than 300 different patterns of spatially-fixed 3Q systems, 
we calculate the 3Q potential $V_{\rm 3Q}$ 
from the 3Q Wilson loop $W_{\rm 3Q}$ 
using SU(3) lattice QCD\cite{TS0102,OST05} 
with the standard plaquette action at the quenched level 
on various lattices, i.e.,  
($\beta$=5.7, $12^3\times 24$),
($\beta$=5.8, $16^3\times 32$), 
($\beta$=6.0, $16^3\times 32$) and 
($\beta=6.2$, $24^4$).
For the accurate measurement, we construct the ground-state-dominant  
3Q operator using the smearing method.
Note also that the lattice QCD calculation is completely independent of any Ansatz for the potential form.

To conclude, 
we find that the static 3Q potential $V_{\rm 3Q}$
is well described by the Coulomb plus Y-type linear potential (Y-Ansatz)  
within 1\%-level deviation.\cite{TS0102}
We also find the universality of the string tension, $\sigma_{\rm 3Q} \simeq \sigma_{\rm Q\bar Q}$, and 
the OGE result, $A_{\rm 3Q} \simeq A_{\rm Q\bar Q}/2$.
As an example, 
we show in Fig.1(a) the 3Q confinement potential $V_{\rm 3Q}^{\rm conf}$, 
i.e., the 3Q potential subtracted by the Coulomb part, 
plotted against the Y-shaped flux-tube length $L_{\rm min}$.
At each $\beta$, clear linear correspondence is found between 
$V_{\rm 3Q}^{\rm conf}$ and $L_{\rm min}$, 
which indicates Y-Ansatz for the 3Q potential.\cite{OST05} 

\begin{figure}[h]
\vspace{-0.3cm}
\begin{center}
\includegraphics[height=4cm]{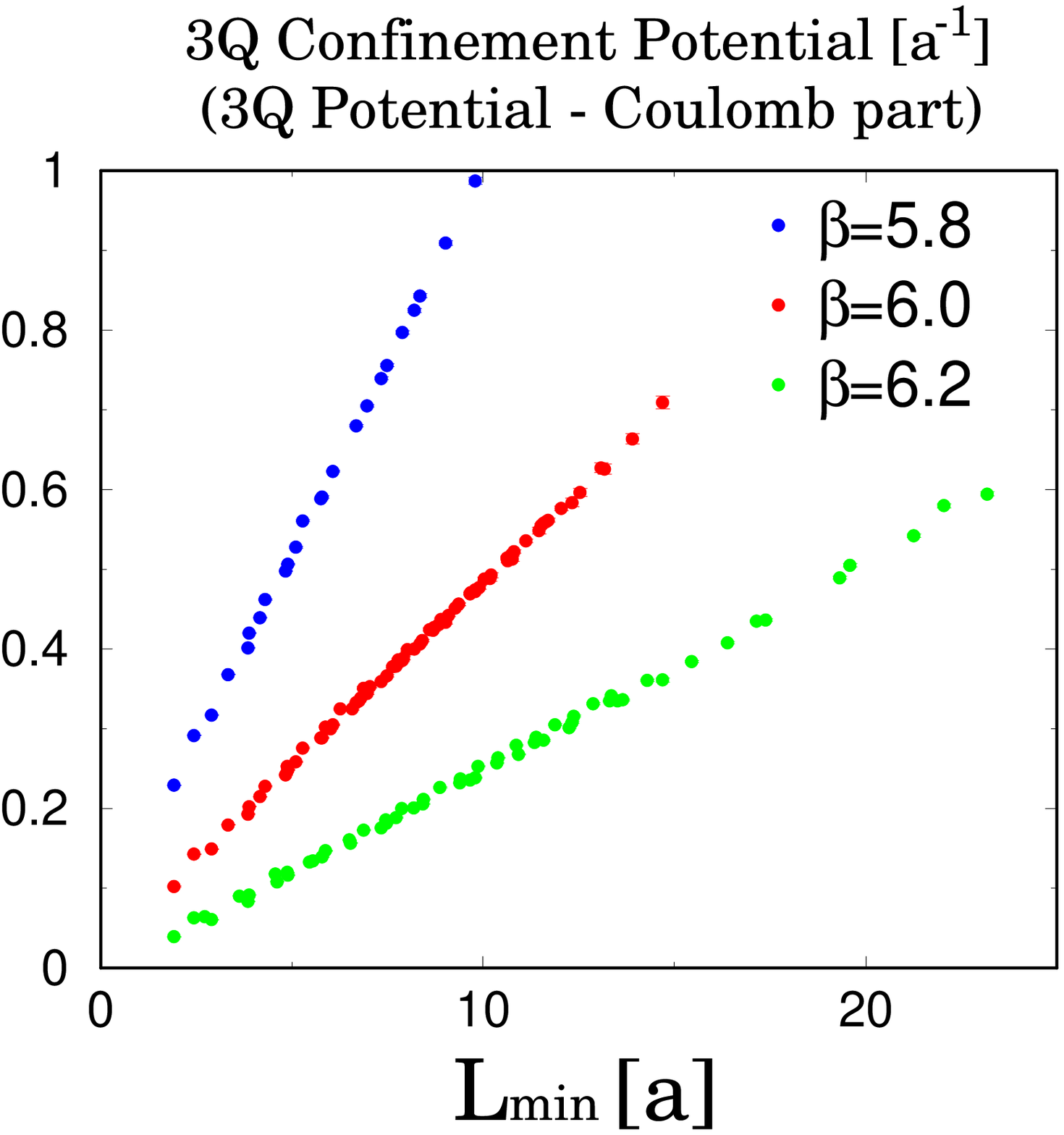}
\includegraphics[height=3.2cm]{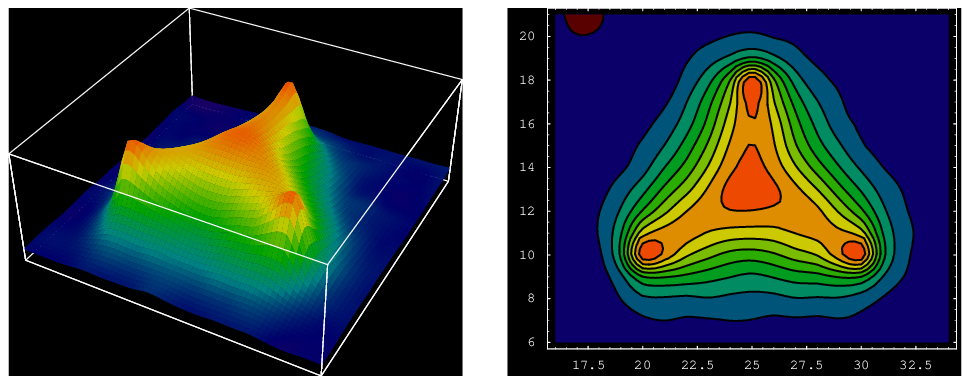}
\caption{
(a) (Left) The 3Q confinement potential $V_{\rm 3Q}^{\rm conf}$, 
i.e., the 3Q potential subtracted by the Coulomb part, 
plotted against 
the total flux-tube length $L_{\rm min}$ of Y-Ansatz
in the lattice unit.$^{10}$
(b) (Middle and right) Lattice results for Y-type flux-tube formation 
in the spatially-fixed 3Q system 
in MA projected QCD.
The distance between the junction and each quark is about 0.5fm.$^{11}$
}
\end{center}
\vspace{-0.7cm}
\end{figure}

\subsection{Quark Confinement Mechanism in Baryons}

As another clear evidence for Y-Ansatz, 
Y-type flux-tube formation is actually observed 
in maximally-Abelian (MA) projected lattice QCD 
from the measurement of the action density  
in the spatially-fixed 3Q system,\cite{Ichie03} as shown in Fig.1(b).

MA projected QCD includes electric and magnetic currents, 
which can be separated as the photon part and the monopole part with the Hodge decomposition.\cite{Ichie03,IS9900} 
We investigate these parts of the 3Q potential in MA projected QCD 
in quenched SU(3) lattice QCD with $16^4$ and $\beta$=6.0, and find the following result.
\begin{itemlist}
\item[1.]
The monopole part $V_{\rm 3Q}^{\rm Mo}$ of the 3Q potential is almost single-valued function of $L_{\rm min}$, 
and is approximated as $V_{\rm 3Q}^{\rm Mo} \simeq \sigma_{\rm 3Q} L_{\rm min}$ at long distances.
\item[2.]
The photon part $V_{\rm 3Q}^{\rm Ph}$ of the 3Q potential is almost single-valued function of 
${L_{\rm Coul}} \equiv (\frac{1}{a}+\frac{1}{b}+\frac{1}{c})^{-1}$ with $a$, $b$ and $c$ being the three sides of the 3Q triangle.
\end{itemlist}
Thus, whereas the electric current induces Coulomb forces, 
the magnetic-monopole current induces the Y-type linear confinement potential for quarks in baryons.

In this way, together with recent several analytical studies\cite{KS03,C0405} 
and other recent lattice QCD studies, Y-Ansatz for the static 3Q potential is almost settled. 

\begin{figure}
\centering
\rotatebox{-90}{\includegraphics[width=3.6cm,clip]{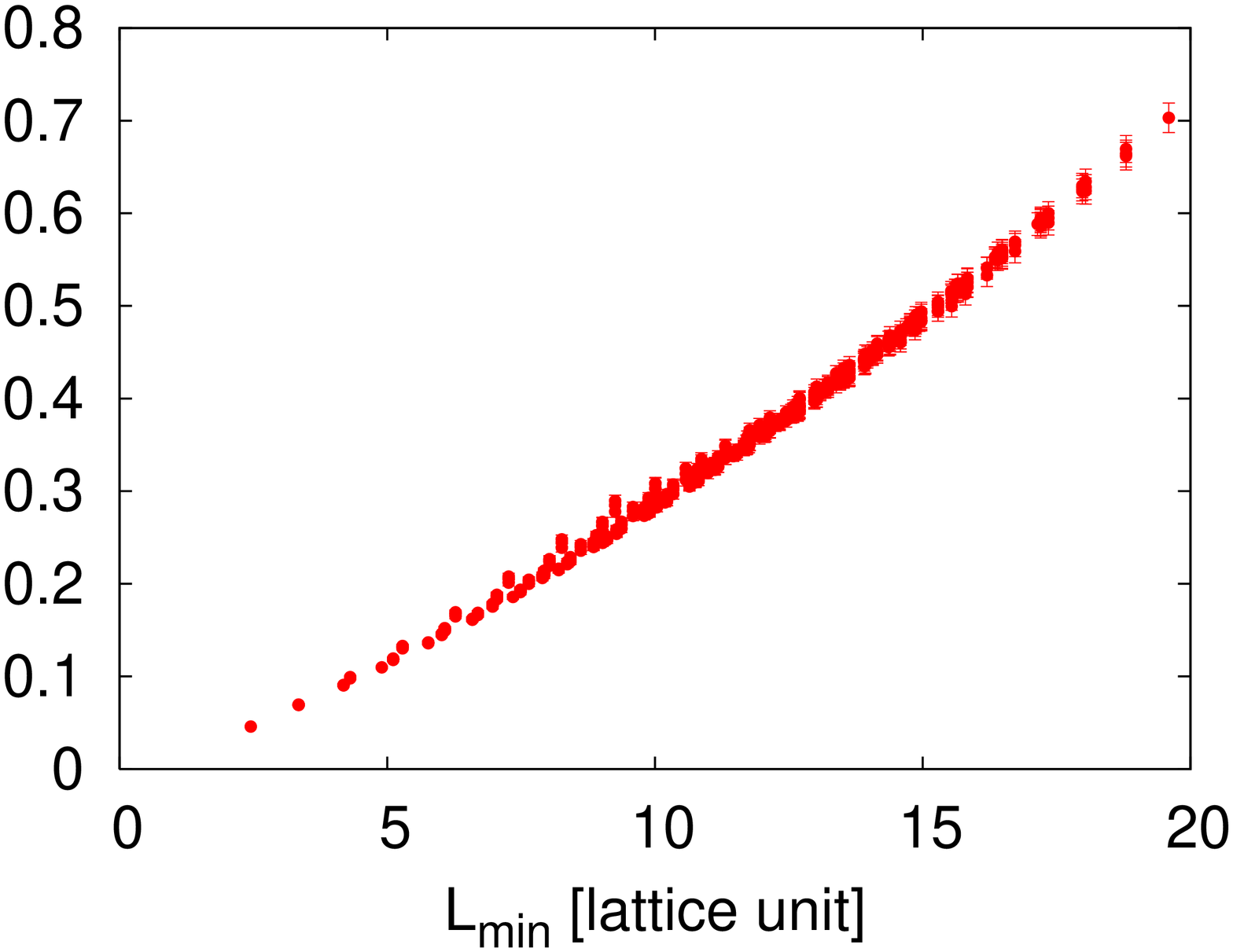}}
\rotatebox{-90}{\includegraphics[width=3.6cm,clip]{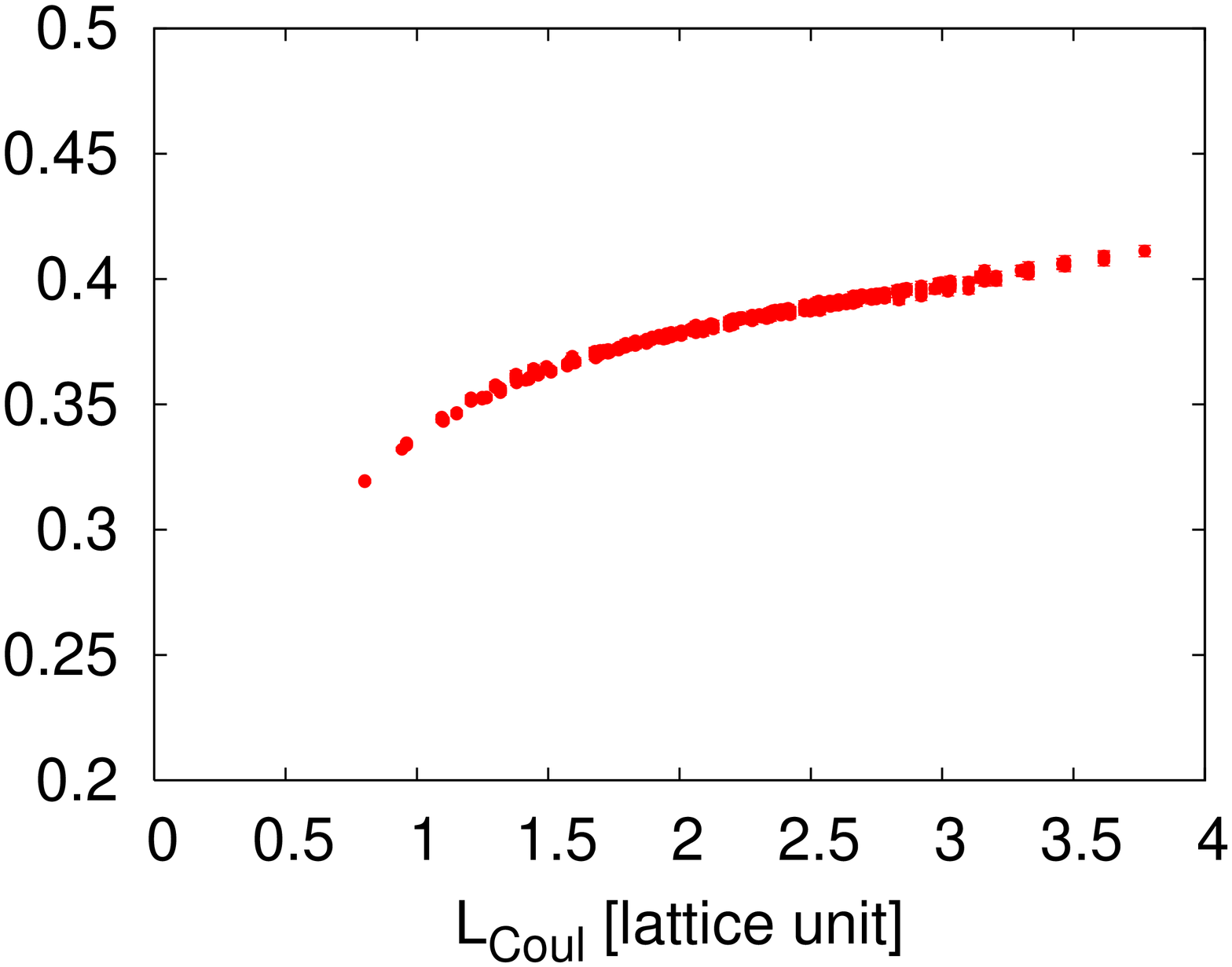}}
\caption{The lattice results for Hodge decomposition of the 3Q potential in MA projected QCD: 
(a) The monopole part $V_{\rm 3Q}^{\rm Mo}$ plotted against $L_{\rm min}$;
(b) the photon part $V_{\rm 3Q}^{\rm Ph}$ plotted against $L_{\rm Coul}$.
}
\end{figure}

\section{Inter-Quark Potentials in Multi-Quark Systems in Lattice QCD}

Next, we perform the first study of 
the inter-quark interaction in multi-quark systems in SU(3) lattice QCD.\cite{OST05}
As for the potential at short distances, 
the perturbative OGE potential would be appropriate, 
due to the asymptotic nature of QCD. 
For the long-range part, however, there appears the confinement potential as a typical 
nonperturbative property of QCD, and its form is highly nontrivial 
in the multi-quark system. 
In fact, to clarify the confinement force in multi-quark systems 
is one of the essential points for the 
construction of the QCD-based quark-model Hamiltonian.
Then, we investigate the multi-quark potential in lattice QCD, 
with paying attention to the confinement force in multi-quark hadrons.

\subsection{One-Gluon-Exchange (OGE) Coulomb plus Multi-Y Ansatz}

We first consider the theoretical form of the multi-quark potential, 
since we will have to analyze the lattice QCD data by comparing them 
with some theoretical Ansatz.
By generalizing the lattice QCD result of Y-Ansatz for the three-quark potential, 
we propose the OGE Coulomb plus multi-Y Ansatz,\cite{OST05}
\begin{eqnarray}
V_{n{\rm Q}}=\frac{3}{2}A_{n{\rm Q}} \sum_{i<j}\frac{T^a_i T^a_j}{|{\bf r}_i-{\bf r}_j|}+\sigma_{n{\rm Q}} L_{\rm min}+C_{n{\rm Q}}~~~~ (n=3,4,5,...),
\label{VnQ}
\end{eqnarray}
for the potential form of the multi-quark system.
Here, the confinement potential is proportional to the minimal total length $L_{\rm min}$ 
of the color flux tube linking the quarks, which is multi-Y shaped in most cases.

In the following, we study the inter-quark interaction in multi-quark systems 
in lattice QCD, and compare the lattice data with the theoretical Ansatz (\ref{VnQ}).
Note that the lattice QCD data themselves are meaningful as primary data 
on the multi-quark system directly based on QCD, and do not depend on any theoretical Ansatz.

\subsection{Formalism of the Multi-Quark Wilson Loop}

We formulate the multi-quark Wilson loop to obtain 
the multi-quark potential in lattice QCD.\cite{OST05}
Similar to the derivation of the Q$\rm\bar{Q}$ potential from the Wilson loop, 
the static multi-quark potential can be derived from 
the corresponding multi-quark Wilson loop.  
We construct the tetraquark Wilson loop $W_{\rm 4Q}$ and the pentaquark Wilson loop $W_{\rm 5Q}$ in a gauge invariant way  
as shown in Fig.3, and define them as 
\begin{eqnarray}
W_{\rm 4Q}&\equiv& \frac{1}{3}{\rm tr}(\tilde{M}_1 \tilde{R}_{12} \tilde{M}_2 \tilde{L}_{12}), \\
W_{\rm 5Q}&\equiv& \frac1{3!}\epsilon^{abc}\epsilon^{a'b'c'}M^{aa'}(\tilde R_3\tilde R_{12}\tilde R_4)^{bb'}(\tilde L_3\tilde L_{12}\tilde L_4)^{cc'}, 
\end{eqnarray}
where $\tilde{M}$, $\tilde{M}_i$, $\tilde{L_j}$, $\tilde{R_j}$ ($i$=1,2, $j$=1,2,3,4), $\tilde{R}_{12}$ and $\tilde{L}_{12}$ are given by 
\begin{eqnarray}
\tilde{M}, \tilde{M}_i, \tilde{R_j}, \tilde{L_j}
\equiv 
P \exp\{ig \int_{M, M_i,R_j,L_j}dx^\mu A_\mu (x)\}
\in \rm{SU(3)_{\rm c}}, 
\\
\tilde{R}_{12}^{a'a} 
\equiv \frac{1}{2}\epsilon^{abc}\epsilon^{a'b'c'}
R_1^{bb'}R_2^{cc'},\quad 
\tilde{L}_{12}^{a'a} 
\equiv \frac{1}{2}\epsilon^{abc}\epsilon^{a'b'c'}
L_1^{bb'}L_2^{cc'}.
\end{eqnarray}
The multi-quark Wilson loop physically means that 
a gauge-invariant multi-quark state is generated at $t=0$ and annihilated at $t=T$ 
with quarks being spatially fixed in ${\bf R}^3$ for $0<t<T$.
The multi-quark potential is obtained from the vacuum expectation value of 
the multi-quark Wilson loop as
\begin{eqnarray}
V_{\rm 4Q}=-\lim_{T\rightarrow \infty} \frac1{T} {\rm ln} \langle W_{\rm 4Q}\rangle, 
\quad
V_{\rm 5Q}=-\lim_{T\rightarrow \infty} \frac1{T} {\rm ln} \langle W_{\rm 5Q}\rangle.
\end{eqnarray}

\begin{figure}[h]
\vspace{-0.3cm}
\centering
\includegraphics[scale=0.33,clip]{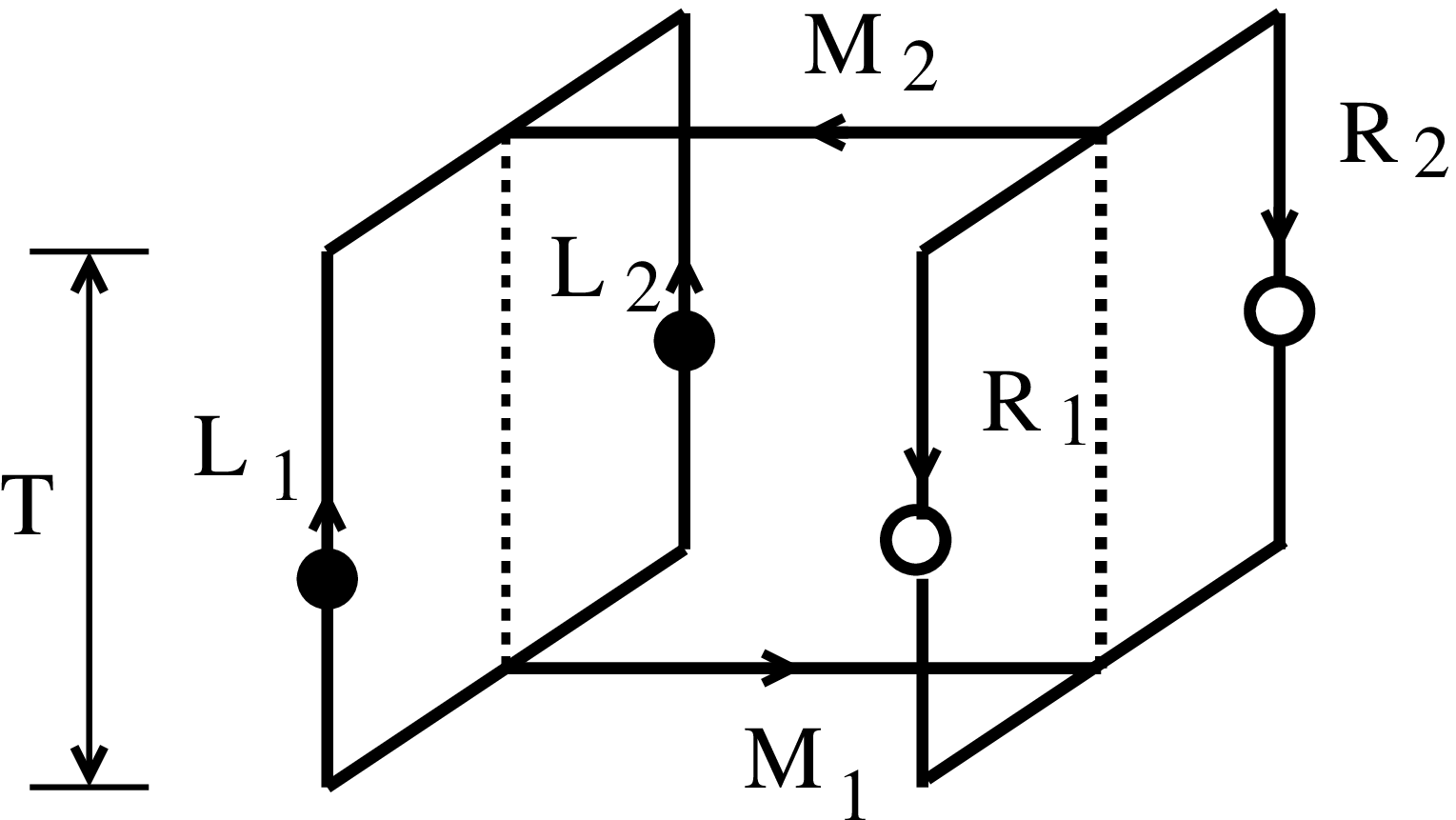}
\includegraphics[scale=0.33,clip]{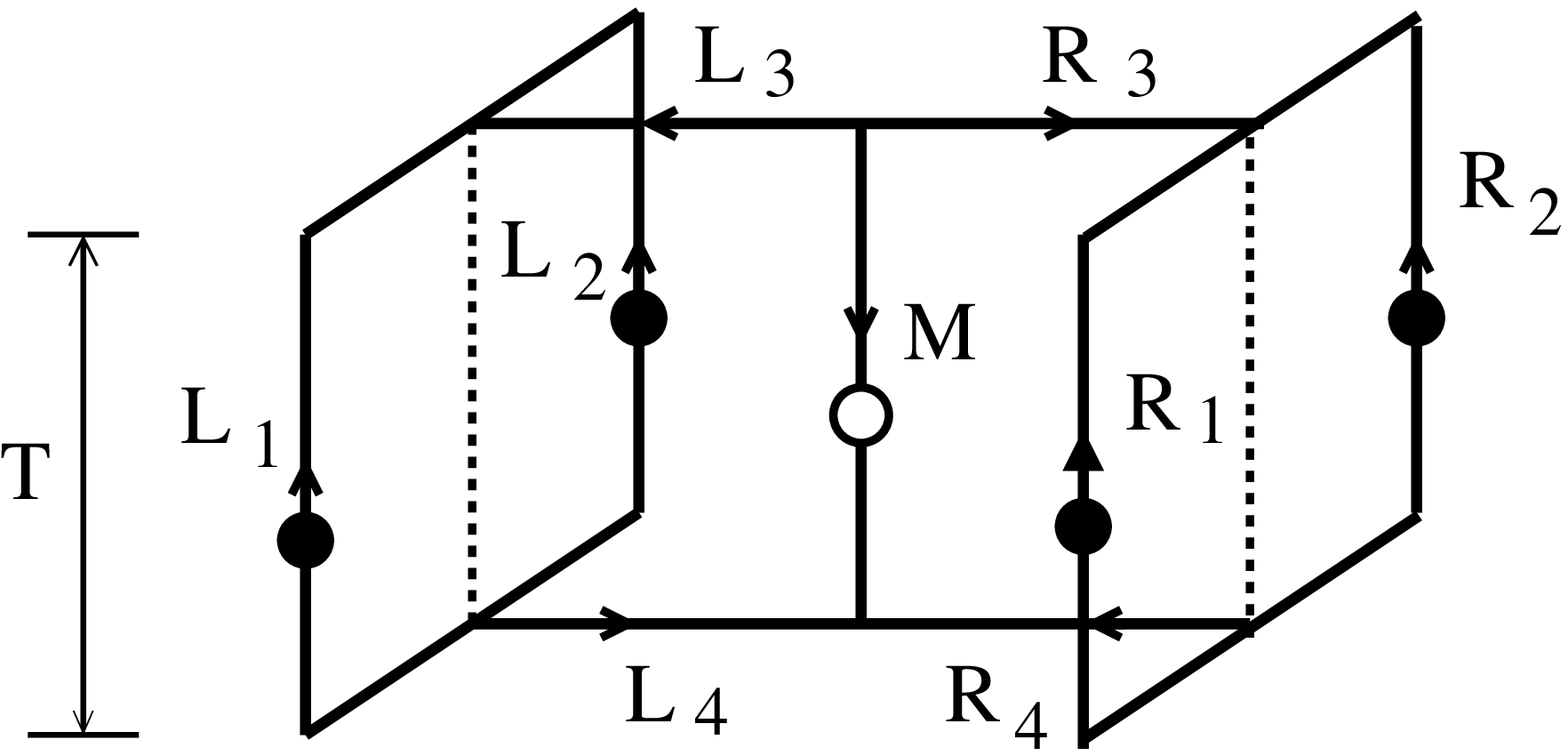}
\caption{(a) The tetraquark Wilson loop $W_{\rm 4Q}$. (b) The pentaquark Wilson loop $W_{\rm 5Q}$. 
The contours $M,M_i,R_j,L_j (i=1,2,j=3,4)$ are line-like and $R_j,L_j (j=1,2)$ are staple-like. 
}
\vspace{-0.6cm}
\end{figure}

\subsection{Tetraquark Potential and Flip-Flop in SU(3) Lattice QCD}

For about 200 different patterns of QQ-${\rm \bar{Q}\bar{Q}}$ configurations, i.e., tetraquark (4Q) systems, 
we perform the detailed study of the tetraquark (4Q) potential $V_{\rm 4Q}$ 
in SU(3) lattice QCD with $\beta$=6.0 and $16^3\times 32$,
and find the following results.\cite{OST05}
\begin{itemlist}
\item[1.]
When QQ and $\rm \bar Q \bar Q$ are well separated,  
the 4Q potential $V_{\rm 4Q}$ is well described by the OGE Coulomb plus multi-Y Ansatz,
which indicates the formation of the multi-Y-shaped flux-tube connecting four (anti)quarks as shown in Fig.4(a).
\item[2.]
When the nearest quark-antiquark pair is spatially close, 
$V_{\rm 4Q}$ is well described by the sum of two Q$\bar {\rm Q}$ potentials, 
which indicates a ``two-meson" state as Fig.4(b).
\end{itemlist}

\begin{figure}[h]
\begin{center}
\includegraphics[height=2.7cm]{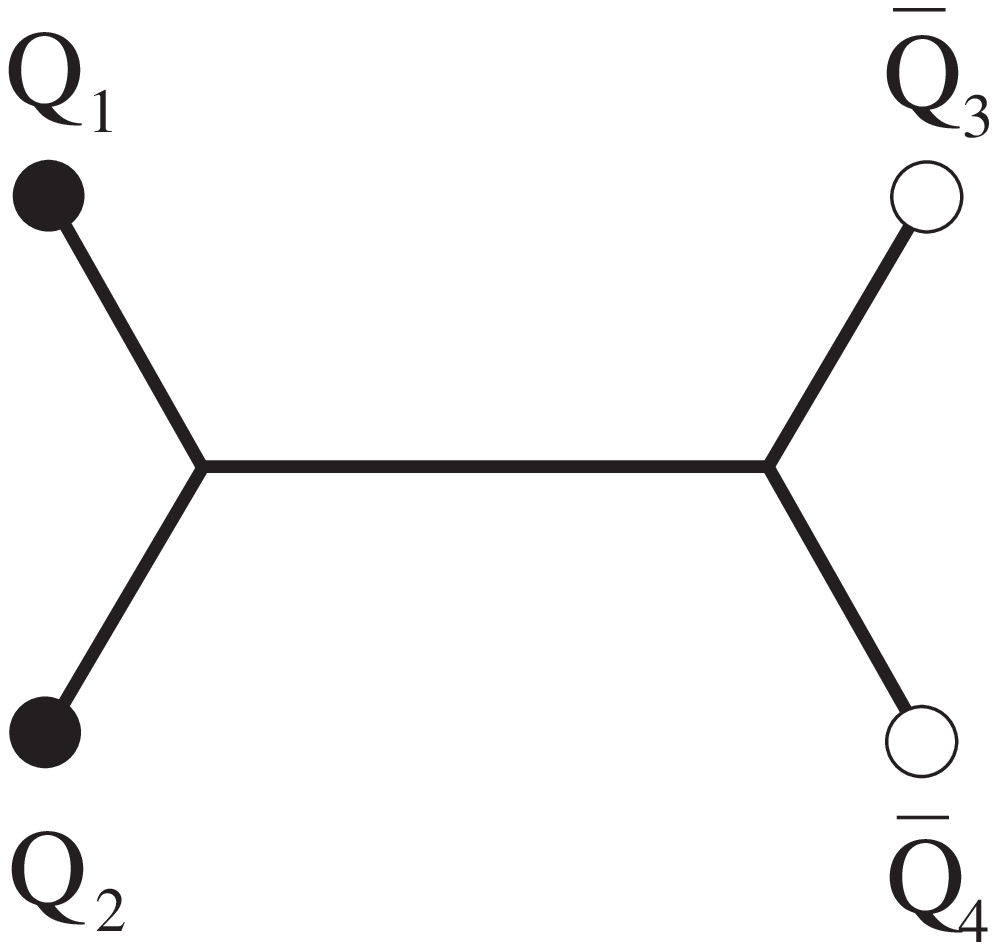}
\hspace{0.8cm}
\includegraphics[height=2.7cm]{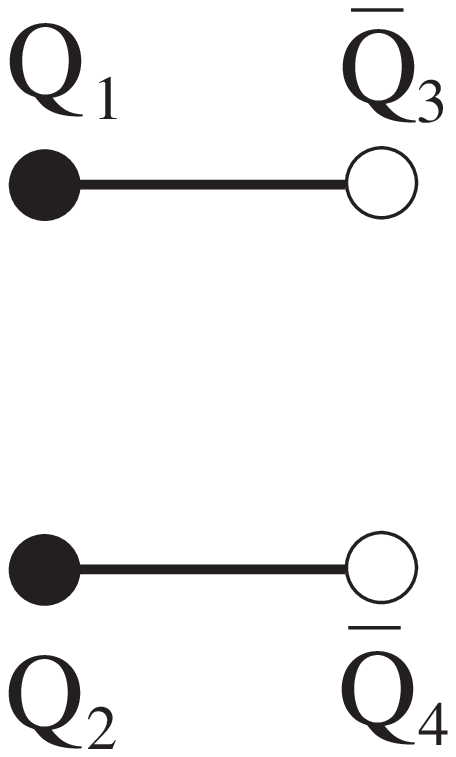} 
\hspace{0.8cm}
\includegraphics[height=2.7cm]{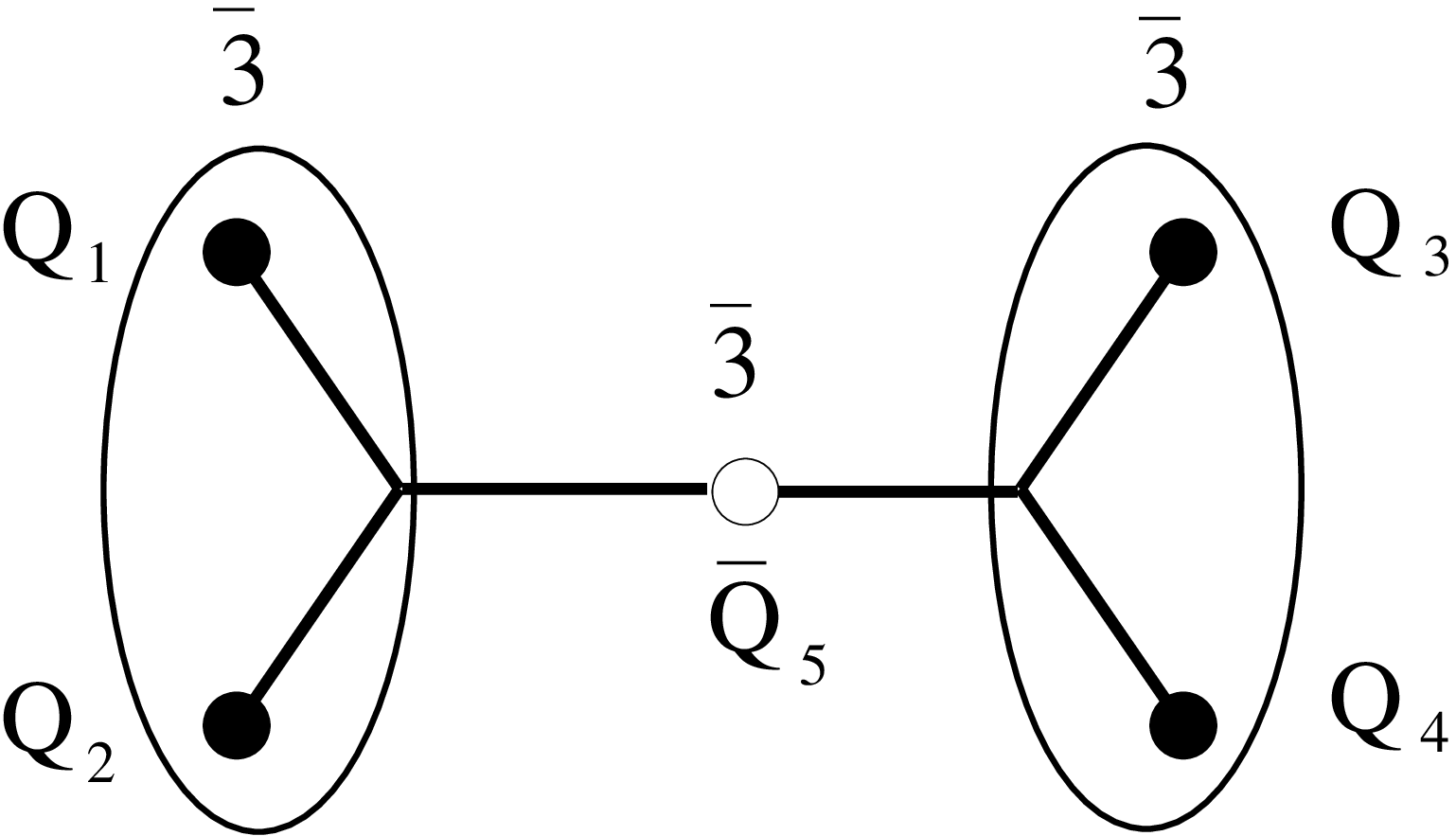} 
\caption{
(a) A connected tetraquark (QQ-$\rm \bar Q\bar Q$) configuration 
and (b) a ``two-meson" configuration.
(c)
A QQ-$\rm \bar Q$-QQ type pentaquark configuration.
The lattice QCD results indicate the multi-Y-shaped flux-tube formation 
for the connected 4Q system and the QQ-$\rm \bar Q$-QQ system.
}
\end{center}
\vspace{-0.2cm}
\end{figure}

As the examples, 
we show in Fig.5 the lattice QCD results of the 4Q potential $V_{\rm 4Q}$ 
for symmetric planar 4Q configurations as shown in Figs.4(a) and (b), 
where each 4Q system is labeled by 
$d\equiv \overline{{\rm Q}_1{\rm Q}_2}/2$ and $h\equiv 
\overline{{\rm Q}_1{\rm Q}_3}$.
For large values of $h$ compared with $d$, the lattice data obey  
the OGE Coulomb plus multi-Y Ansatz.
For small $h$, the lattice data obey the ``two-meson" Ansatz, 
where the 4Q potential is described by the sum of two Q$\bar {\rm Q}$ potentials, 
$V_{\rm Q\bar Q}(r_{13})+V_{\rm Q\bar Q}(r_{24})=2V_{\rm Q\bar Q}(h)$. 

Thus, the tetraquark potential $V_{\rm 4Q}$ is found to take 
the smaller energy of the connected 4Q state or the two-meson state.\cite{OST05}
In other words, we observe a clear lattice QCD evidence 
of the ``flip-flop", i.e., the flux-tube recombination 
between the connected 4Q state and the two-meson state.

\begin{figure}[h]
\vspace{-0.2cm}
\centering
\includegraphics[height=3.8cm,clip]{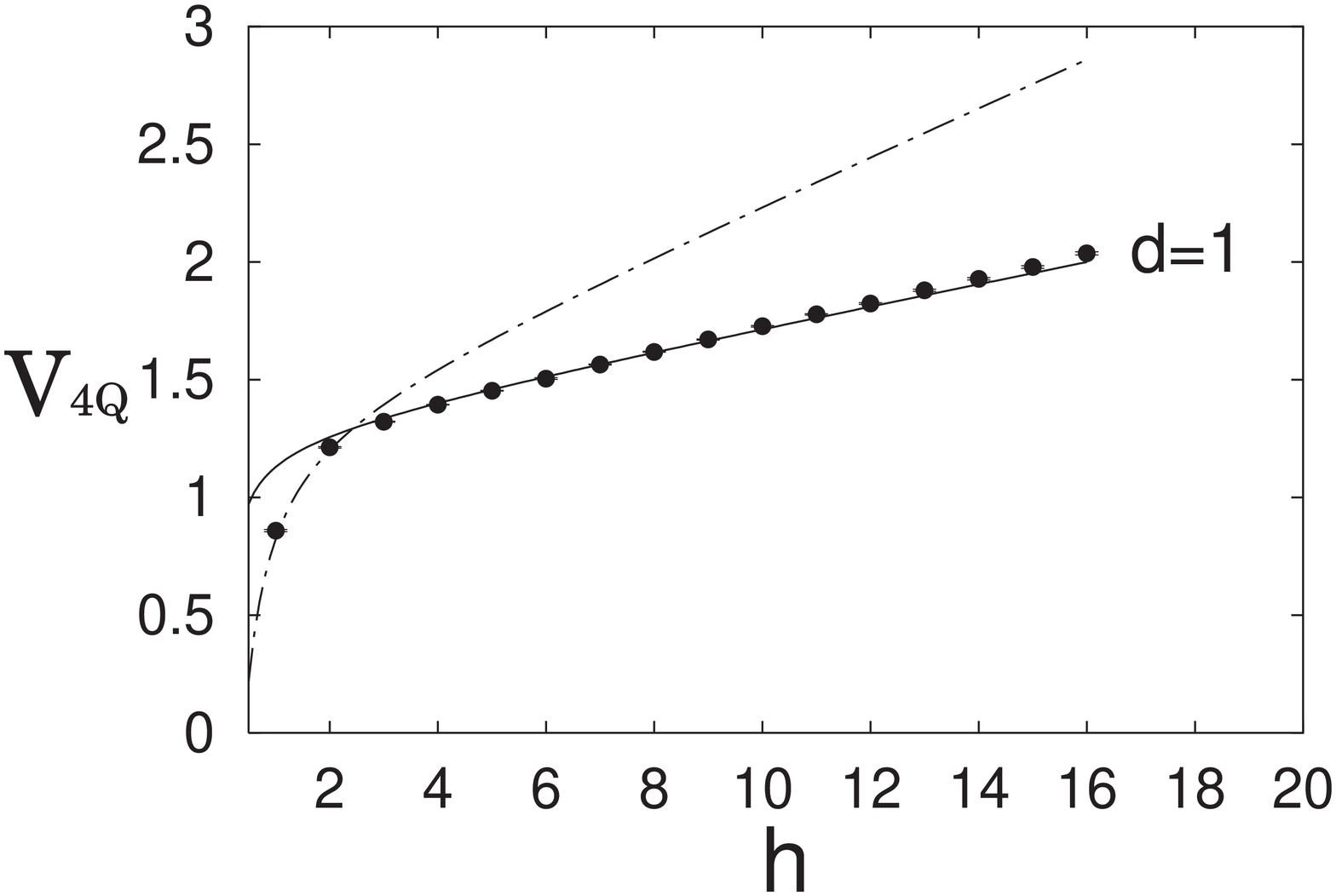}
\hspace{0.1cm}
\includegraphics[height=3.8cm,clip]{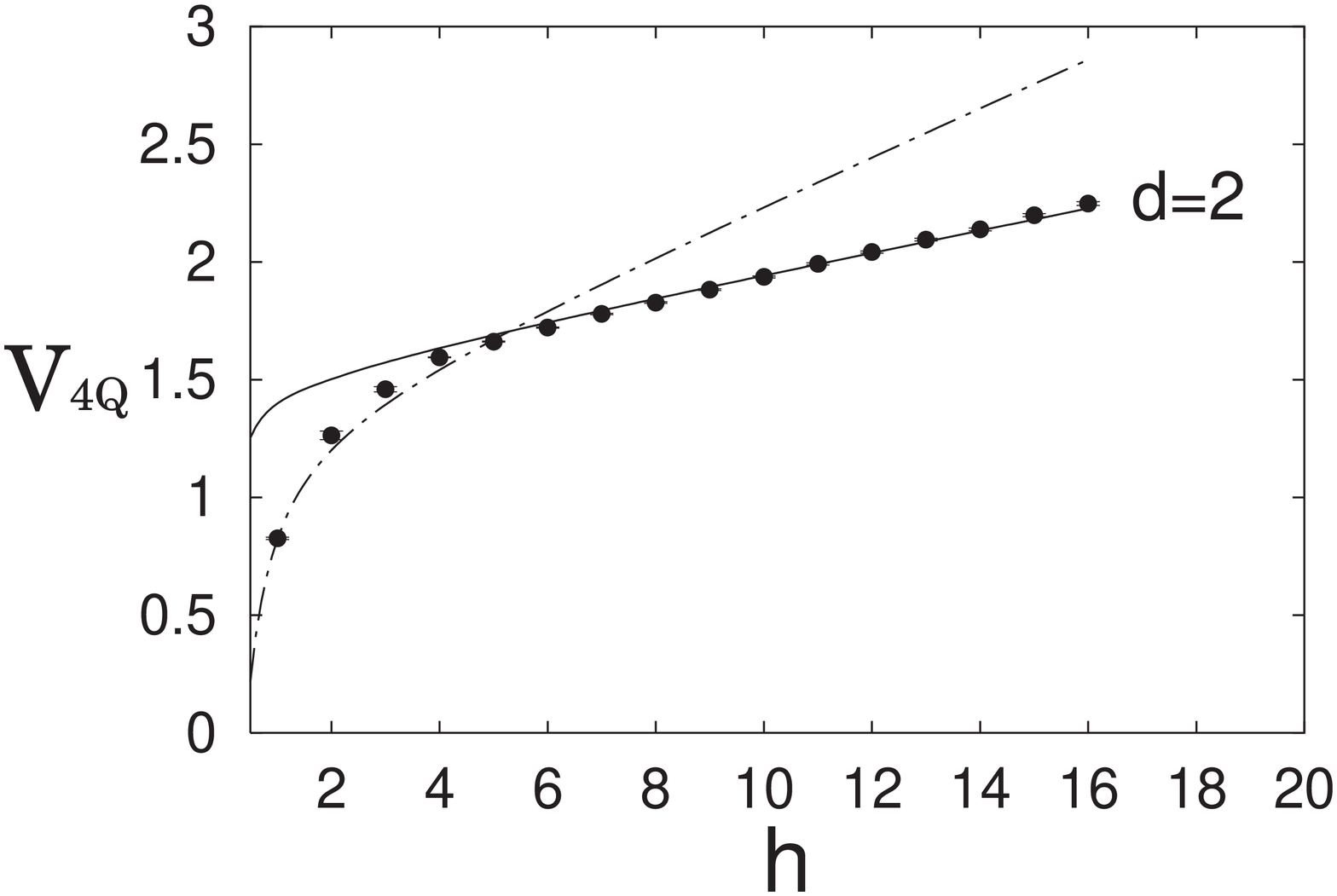}
\caption{
Lattice QCD results of the tetraquark potential $V_{\rm 4Q}$ 
for symmetric planar 4Q configurations in the lattice unit.
The symbols denote the lattice QCD data. 
The solid curve denotes the OGE plus multi-Y Ansatz, 
and the dotted-dashed curve the two-meson Ansatz.}
\vspace{-0.7cm}
\end{figure}

\subsection{Lattice QCD Result of the Pentaquark Potential}

We perform the first study of the pentaquark (5Q) potential $V_{\rm 5Q}$ in lattice QCD with $\beta$=6.0 and $16^3\times 32$
for 56 different patterns of QQ-$\rm \bar Q$-QQ type pentaquark configurations.\cite{OST05}
We find that the lattice QCD data of  
$V_{\rm 5Q}$ are well described by the OGE Coulomb plus multi-Y Ansatz, i.e., 
the sum of the OGE Coulomb term and the multi-Y-type 
linear term based on the flux-tube picture.

We show in Fig.6 the lattice QCD results of the 5Q potential $V_{\rm 5Q}$ 
for symmetric planar 5Q configurations as shown in Fig.4(c), 
where each 5Q system is labeled by 
$d\equiv \overline{{\rm Q}_1{\rm Q}_2}/2$ and $h\equiv 
\overline{{\rm Q}_1{\rm Q}_3}$.
In Fig.6, we add the theoretical curves of  
the OGE Coulomb plus multi-Y Ansatz, where the coefficients 
$(A_{\rm 5Q},\sigma_{\rm 5Q})$ are set to be 
$(A_{\rm 3Q},\sigma_{\rm 3Q})$ obtained from the 3Q potential.\cite{TS0102}
In Fig.6, one finds a good agreement between 
the lattice QCD data of $V_{\rm 5Q}$ and the theoretical curves.

In this way, the multi-quark potentials $V_{n{\rm Q}}$ ($n$=3,4,5) 
are found to be well described by the OGE Coulomb plus multi-Y Ansatz.\cite{TS0102,OST05} 
These lattice QCD results support the flux-tube picture even for the multi-quark systems.

\begin{figure}[h]
\vspace{-0.25cm}
\begin{center}
\includegraphics[height=3.6cm]{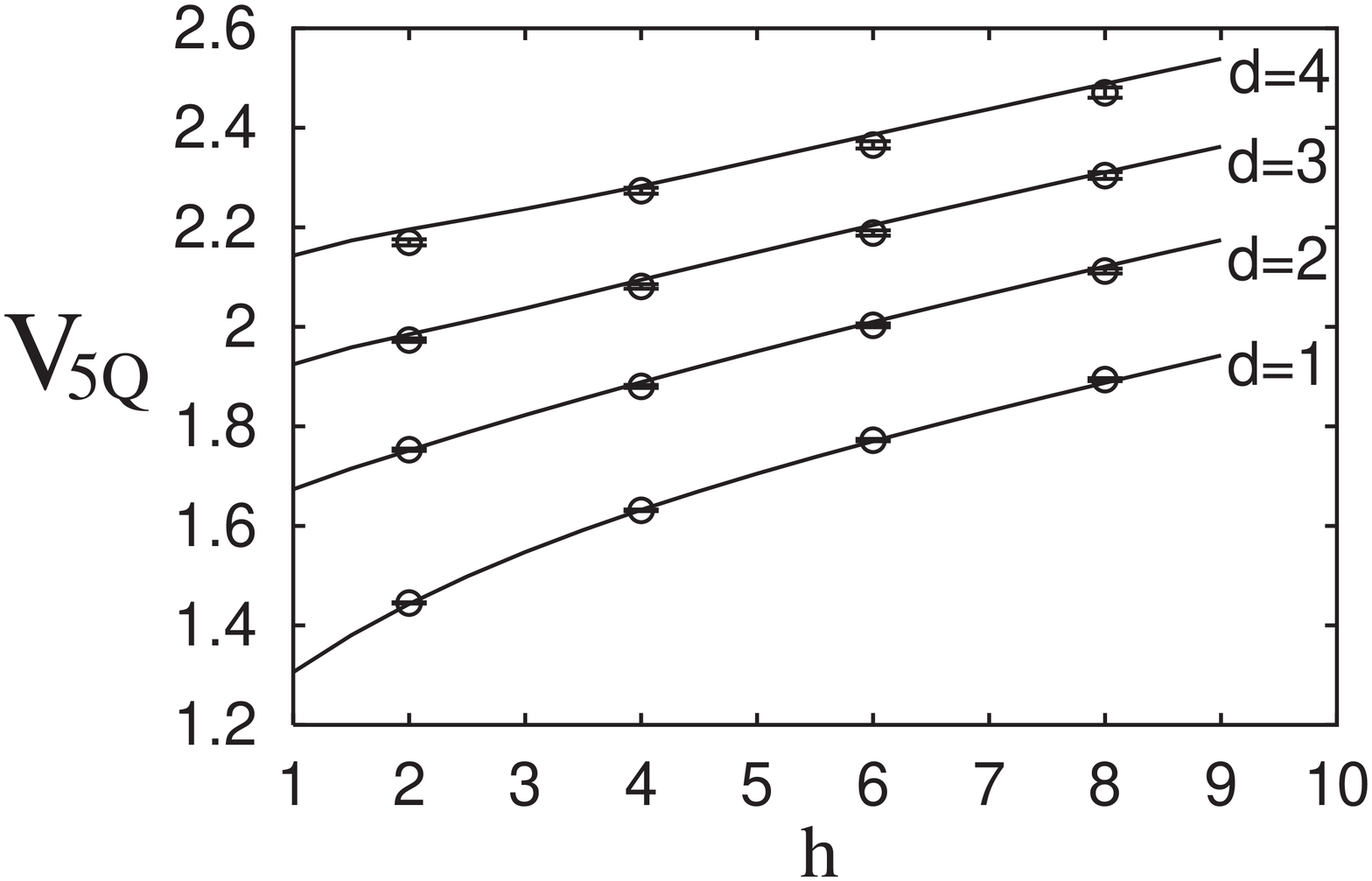}
\includegraphics[height=3.6cm]{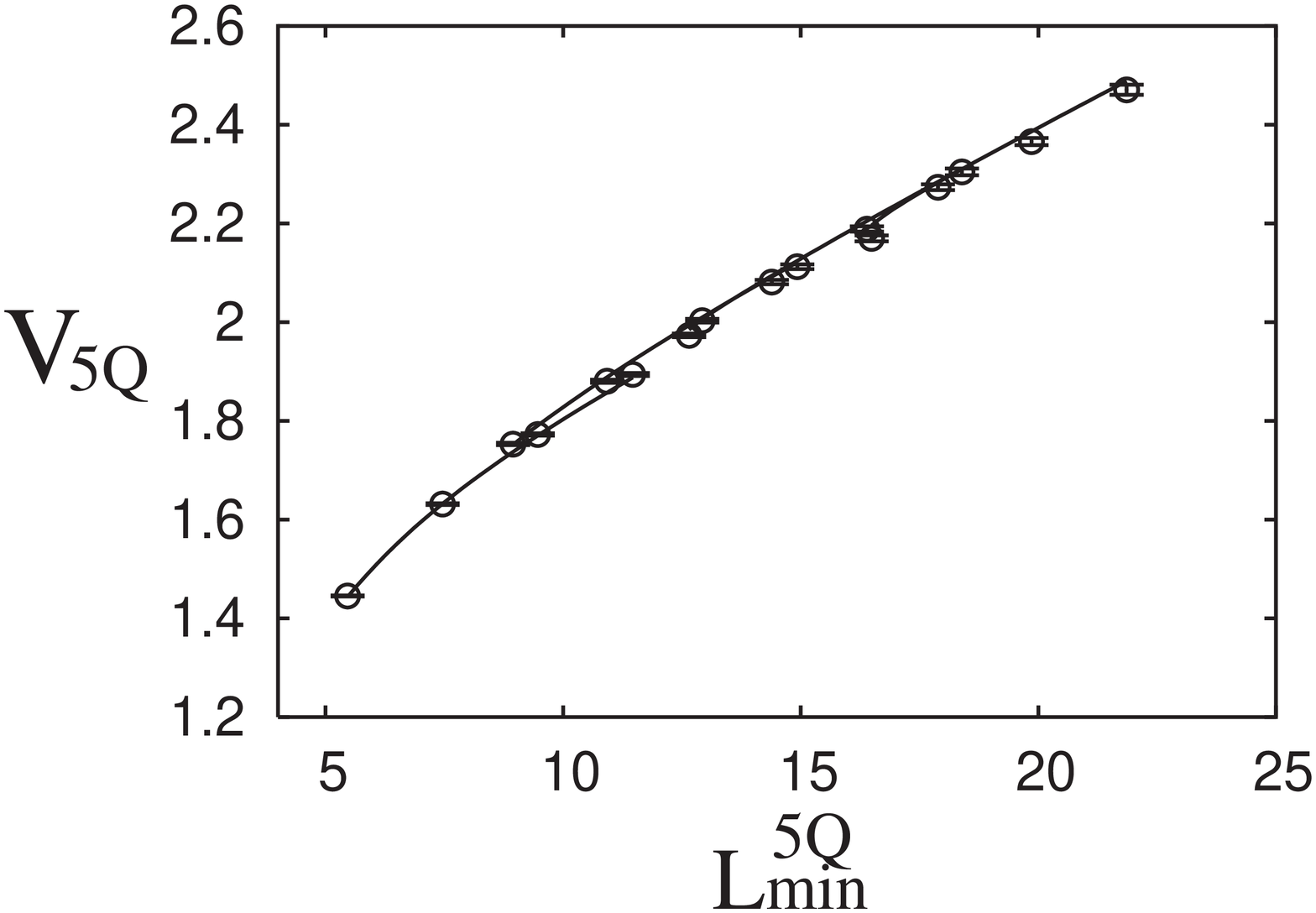}
\caption{
Lattice QCD results of the pentaquark potential $V_{\rm 5Q}$ 
for symmetric planar 5Q configurations as shown in Fig.4(c) in the lattice unit:  
(a) $V_{\rm 5Q}$ v.s. $(d,h)$ and (b) $V_{\rm 5Q}$ v.s. $L_{\rm min}^{\rm 5Q}$.
The symbols denote the lattice QCD data, and the curves the theoretical form 
of the OGE plus multi-Y Ansatz.
The lattice QCD results indicate the multi-Y-shaped flux-tube formation 
in the QQ-$\rm \bar Q$-QQ system.
}
\end{center}
\vspace{-0.8cm}
\end{figure}

\section{Heavy-Heavy-Light Quark Potential and Quark Motional Effect}

To see quark motional effects in baryons, we perform the first study of 
the heavy-heavy-light quark (QQq) potential both in lattice QCD\cite{YSI07} and in a quark model.\cite{YS08}

We calculate the QQq potential $V_{\rm QQq}(R)$ as the function of the distance $R$ 
between the two heavy quarks in SU(3) quenched lattice QCD with ($16^4$, $\beta=6.0$) 
and $O(a)$-improved Wilson fermions.\cite{YSI07} 
$V_{\rm QQq}(R)$ is found to be well described with a Coulomb plus linear potential form, 
$V_{\rm QQq}(R)=-\frac{A_{\rm eff}}{R}+\sigma_{\rm eff}R+C_{\rm eff}$, for $R \le$ 0.8fm.
We find that, compared with the static three-quark case, 
the effective confining force $\sigma_{\rm eff}$ between the two heavy quarks is 
reduced in the QQq system by the quark motional effect, e.g., $\sigma_{\rm eff} \simeq 0.8 \sigma_{\rm 3Q}$ 
for the constituent quark mass $M_q \simeq$ 0.5GeV.\cite{YSI07}

We also calculate the QQq potential $V_{\rm QQq}(R)$ in the quark potential model 
with the three-quark confinement potential obtained by lattice QCD.\cite{YS08} 
The light-quark wave-function $\psi(\vec{r})$ distributes in the spatial region 
between the two heavy quarks QQ, as shown in Fig.7. 
We find again reduction of the effective confining force $\sigma_{\rm eff}$ between QQ in the QQq system, 
e.g., $\sigma_{\rm eff} \simeq 0.8 \sigma_{\rm 3Q}$ for $M_q \simeq$ 0.5GeV.\cite{YS08}

\begin{figure}[h]
\begin{center}
\includegraphics[height=3.6cm]{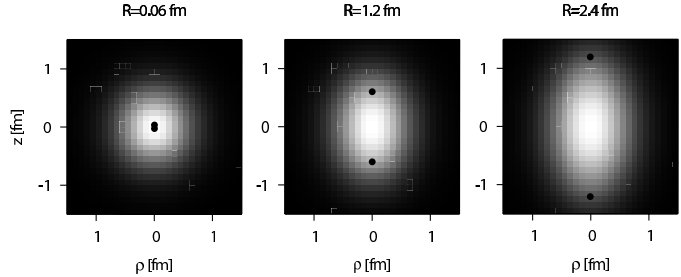}
\caption{
The light-quark spatial distribution $|\psi (\vec{r})|^2$ for 
$M_q$=330MeV with $R$=0.06fm (left), $R$=1.2fm (center) and $R$=2.4fm (right) in QQq systems.
The brighter region has higher probability, and the black circles denote the positions of the heavy quarks.
}
\end{center}
\end{figure}

Both in lattice QCD and in a quark model, 
the inter-two-quark confining force is effectively reduced by the motional effect of 
the remaining ``3rd''quark in baryons.
We conjecture that the effective reduction of the inter-two-quark confining force 
generally occurs also in light-quark baryons due to the 3rd quark motional effect.

\section{Summary and Concluding Remarks}

We have performed the first detailed studies of various inter-quark potentials in SU(3) lattice QCD.
The static three- and multi-quark potentials 
are well described by Y-Ansatz and the OGE Coulomb plus multi-Y type linear potential, respectively. 
This supports the flux-tube picture for baryons and multi-quark hadrons. 
We have also studied heavy-heavy-light (QQq) quark systems, and 
have found the quark motional effect to reduce effectively the inter-two-quark confining force in baryons.
These lattice QCD results are useful to construct the QCD-based quark model.


\begin{thebibliography}{0}
\bibitem{N66} Y.~Nambu, in {\it Preludes in Theoretical Physics}, 
(North-Holland, 1966).
\bibitem{HN65} M.~Y.~Han and Y.~Nambu, {\it Phys. Rev.} {\bf 139}, B1006 (1965). 
\bibitem{GWP73}
D.~J.~Gross and F.~Wilczek, {\it Phys. Rev. Lett.} {\bf 30}, 1343 (1973);
H.~D.~Politzer, {\it Phys. Rev. Lett.} {\bf 30}, 1346 (1973).
\bibitem{N6970} Y.~Nambu, in {\it Symmetries and Quark Models} (Wayne State University, 1969).
\bibitem{N74} Y.~Nambu, {\it Phys. Rev. D} {\bf 10}, 4262 (1974).
\bibitem{NJL61} Y.~Nambu and G.~Jona-Lasinio, 
{\it Phys. Rev.} {\bf 122}, 345 (1961); {\it ibid.} {\bf 124}, 246 (1961).
\bibitem{C7980} M.~Creutz, 
{\it Phys. Rev. Lett.} {\bf 43}, 553 (1979); {\it Phys. Rev. D} {\bf 21}, 2308 (1980).
\bibitem{R05} 
H.~J.~Rothe, {\it Lattice Gauge Theories}, 3rd edition (World Scientific, 2005).
\bibitem{TS0102} 
H.~Suganuma, Y.~Nemoto, H.~Matsufuru and T.~T.~Takahashi, {\it Nucl. Phys. A} {\bf 680}, 159 (2000);
T.~T.~Takahashi, H.~Matsufuru, Y.~Nemoto and H.~Suganuma, 
{\it Phys.~Rev.~Lett.} {\bf 86}, 18 (2001);
T.~T.~Takahashi, H.~Suganuma, Y.~Nemoto and H.~Matsufuru, 
{\it Phys.~Rev.~D} {\bf 65}, 114509 (2002);
T.~T.~Takahashi and H.~Suganuma, {\it Phys. Rev. Lett.} {\bf 90}, 182001 (2003); 
{\it Phys. Rev. D} {\bf 70}, 074506 (2004).
\bibitem{OST05}
F.~Okiharu, H.~Suganuma and T.~T.~Takahashi, 
{\it Phys.~Rev.~Lett.} {\bf 94}, 192001 (2005); {\it Phys. Rev. D} {\bf 72}, 014505 (2005).
\bibitem{Ichie03} H.~Ichie, V.~Bornyakov, T.~Streuer and G.~Schierholz, 
{\it Nucl. Phys. A} {\bf 721}, 899 (2003); 
T.~T.~Takahashi, H.~Suganuma, H.~Ichie, H.~Matsufuru, Y.~Nemoto,
{\it Nucl. Phys. A} {\bf 721}, 926 (2003); 
V.~G.~Bornyakov, H.~Ichie, Y.~Mori, D.~Pleiter, M.~I.~Polikarpov, G.~Schierholz, T.~Streuer, H.~St\"uben and T.~Suzuki, 
{\it Phys. Rev. D} {\bf 70}, 054506 (2004);
H.~Suganuma, F.~Okiharu, T.~T.~Takahashi and H.~Ichie, {\it Nucl. Phys. A} {\bf 755}, 399 (2005).
\bibitem{KS03} D.~S.~Kuzmenko and Yu.~A.~Simonov, {\it Phys. Atom. Nucl.} {\bf 66}, 950 (2003).
\bibitem{C0405} J.~M.~Cornwall, {\it Phys. Rev. D} {\bf 69}, 065013 (2004); {\it Phys. Rev. D} {\bf 71}, 056002 (2005).
\bibitem{IS9900} 
H.~Ichie and H.~Suganuma, {\it Nucl. Phys. B} {\bf 548}, 365 (1999); {\it ibid.} {\bf 574}, 70 (2000);
K.~Amemiya and H.~Suganuma, {\it Phys. Rev. D} {\bf 60}, 114509 (1999).
\bibitem{YSI07} A.~Yamamoto, H.~Suganuma and H.~Iida, arXiv:0708.3610 [hep-lat].
\bibitem{YS08} A.~Yamamoto and H.~Suganuma, {\it Phys. Rev. D} {\bf 77}, 014036 (2008).
\end{thebibliography}
\end{document}